\documentclass[prl,twocolumn,lengthcheck,superscriptaddress]{revtex4-1}

\usepackage{amsmath}
\usepackage{amssymb}
\usepackage{graphicx}
\usepackage{hyperref}
\usepackage{color}

\newcommand\id{\mathbb{I}}

\newcommand{\tr}{\operatorname{tr}}

\begin{document}

\title{Holographic quantum states}

\author{Tobias J.\ Osborne}
\affiliation{Wissenschaftskolleg zu Berlin, Berlin 14193, Germany}
\author{Jens Eisert}
\affiliation{Wissenschaftskolleg zu Berlin, Berlin 14193, Germany}
\author{Frank Verstraete}
\affiliation{University of Vienna, Faculty of Physics, Boltzmanngasse 5, A-1090 Wien}

\begin{abstract}
	We show how continuous matrix product states of quantum field theories can be described in terms of the dissipative non-equilibrium dynamics of a lower-dimensional auxiliary boundary field theory. We demonstrate that the spatial correlation functions of the bulk field can be brought into one-to-one correspondence with the temporal statistics of the quantum jumps of the boundary field. This equivalence: (1) illustrates an intimate connection between the theory of continuous quantum measurement and quantum field theory; (2) gives an explicit construction of the boundary field theory allowing the extension of real-space renormalization group methods to arbitrary dimensional quantum field theories without the introduction of a lattice parameter; and (3) yields a novel interpretation of recent cavity QED experiments in terms of quantum field theory, and hence paves the way toward  observing genuine quantum phase transitions in such zero-dimensional driven quantum systems. 
\end{abstract}

\maketitle

In recent years we have witnessed tremendous success in the calculation of physical properties of quantum many-body systems from their wavefunctions. This development has been spurred by studies of the entanglement properties of strongly correlated quantum spin systems: it has been established that natural states of quantum lattice systems  are only slightly entangled, and hence typically obey an entropy area law \cite{verstraete:2005a, osborne:2005d, hastings:2007a,area}. These new developments have also shown that the manifold of physical quantum lattice states are well captured by \emph{matrix product states} (MPS) or \emph{finitely correlated states} (FCS) \cite{fannes:1991a}. 

These developments have also allowed the interpretation of the renormalization methods of Wilson \cite{Wilson} and White \cite{White, schollwoeck:2005a} as applications of the variational principle to matrix product states, and have led to natural generalizations of these RG schemes to higher dimensions \cite{richter:1996, verstraete:2004a, jordan:2008a}. A key property, crucially responsible for the success of these RG schemes, has been the fact that the information concerning the quantum correlations of a natural quantum lattice state is encoded in the variational parameters of an auxiliary \emph{zero-dimensional}  system. 

A natural next step is then to develop a similar approach for quantum field theories, and this is exactly the subject of this paper. However, capturing the manifold of low-energy wavefunctionals in, e.g., bosonic theories is much more challenging due to the continuous infinity of degrees of freedom (as opposed to the lattice of finite-dimensional Hilbert spaces for quantum spin systems). The most natural way to proceed is to discretize the continuous degrees of freedom by imposing a lattice cut-off and truncating the local hilbert spaces at each site \cite{iblisdir:2007a, muth:2009a, weir:2010a}. This is similar to the approach taken by lattice gauge theory which, so far, provides essentially the only systematic way to understand nonperturbative effects in quantum field theories. Recently, however, it was established that there is no need to impose a lattice cutoff because continuum limits of matrix product states can be directly defined, and these states, termed \emph{continuous matrix product states} (cMPS), can represent the low-energy physics of non-relativistic field theories extremely accurately \cite{verstraete:2010a, maruyama:2010a}. 

In this work we describe a method to generate cMPS for quantum field theories. 
Our procedure is based on the paradigm of \emph{continuous measurement} \cite{caves:1987a}, where one subjects a quantum system to a weak sequence of measurements of some physical observable. This procedure is a natural generalisation of the sequential preparation scheme for MPS proposed in \cite{Schon} to the continuous setting and we show that this procedure naturally generates cMPS of Ref.\ \cite{verstraete:2010a} and we relate our approach to that of \cite{iblisdir:2007a}. The perspective offered here also allows one to design more flexible classes of states which can be tailored to a system of interest, including quantum states for bosonic systems at arbitrary filling. Our approach also suggests a natural generalisation to arbitrary dimensions: in the case of a 2+1 dimensional quantum field theory the resulting boundary field evolves according to local dissipative dynamics in 1+1 dimensions. An alternative interpretation of the resulting wavefunctionals is that the boundary field provides a local parameterisation of the bulk field; this realises one of the major prerequisites identified by Feynman for the successful application of the variational principle to quantum field theory \cite{Feynman}.

Surprisingly, the bulk and boundary fields have a direct interpretation in the context of cavity electrodynamics for trapped atoms \cite{Zoller,Kimble,Rempe,Schon}: the role of the auxiliary system is played by the trapped atom and the quantum field describes the photons leaking from the cavity. An atom with a fairly low number of internal addressable levels (e.g. $D=6$), would already allow the reproduction of all static correlations functions in, e.g., the Lieb-Liniger model \cite{Lieb-Liniger}. This is achieved by observing the temporal counting statistics of the photons leaking from the cavity.  The present paper also sheds new light on the recently discovered phase transitions of the quantum trajectories obtained in dissipative systems \cite{Garrahan}: such dynamical phase transitions are in correspondence with static quantum phase transitions of a quantum field theory in one dimension higher. Just as  classical one-dimensional nonequilibrium systems exhibit phase transitions similar to their two-dimensional static analogues, 0+1 dimensional nonequilibrium systems exhibit dynamical phase transitions analogous to quantum phase transitions in 1+1 dimension. 

We begin by modelling a measurement of some physical observable $M$ on a quantum system with dimension $D$, which we initially call the \emph{``system''}. Our model, known as \emph{von Neumann's prescription} \cite{vonneumann:1996a}, 
is defined as follows. We attach a quantum system with a continuous degree of freedom, called the 
\emph{meter}, in a fiducial state vector $|0\rangle$ and 
couple it the system for some time $t$ according to the interaction 
$H_I = M\otimes p$. Supposing the system is initially in $|\phi\rangle$, we see that after the interaction the state of the system and meter is given by
\begin{equation*}
		e^{-itH_I}|\phi\rangle|0\rangle = \sum_{j=1}^D \phi_j |m_j\rangle |x = m_jt\rangle 
\end{equation*}	
where $|m_j\rangle$, $j = 1, 2, \dots, D$, are the eigenstates of $M$ with 
corresponding eigenvalues $m_j$, $|\phi\rangle = \sum_{j=1}^{D} \phi_j|m_j\rangle$, $|x = m_jt\rangle = D(a, \sqrt{2}m_jt)|x = 0\rangle$, and 
$D(a, \alpha) = e^{\alpha a^\dag - \overline{\alpha} a}$ is the (phase-space) displacement operator. In our case, at $t=1$ the displacement operator simply effects a translation of the meter's state, initially localised at position $x=0$, to the locations of the eigenvalues $x=m_j$. It is important to note that here and in the sequel we never actually perform a projective measurement of the meter. This can be thought of as corresponding to the situation where a projective measurement of the meter is performed but the measurement record is discarded. While this is an admittedly rudimentary model of the physical measurement process it does afford a considerable potential for generalisation. 

Now, the core of our proposal is to turn von Neumann's measurement prescription on its head and regard the meter 
(attached as an ancillary system) as the \emph{fundamental system} $\mathcal{A}$ and the system as an \emph{auxiliary ancilla} $\mathcal{B}$. 
In this way we can think of it as a \emph{state generation device}: we can obtain a variety of \emph{physical} quantum 
states of the meter $\mathcal{A}$ alone---a bosonic system with a continuous degree of freedom---by exploiting the measurement 
prescription and then tracing out, or perhaps measuring, the system $\mathcal{B}$. 
In these terms we have a way to generate states of a quantum system with a continuous degree of freedom. The challenge 
remains, however, to somehow exploit this procedure to obtain quantum states of a continuous infinity of such continuous degrees of freedom.  

The way we do this here is to model the \emph{continuous} measurement of a specific POVM (positive operator valued measure \cite{POVMComment}). 
Our model, which follows Ref.\ \cite{caves:1987a} closely, is defined by a family of $D\times D$ \emph{complex} matrices $R(x)$, $x\in [0, L]$
\cite{MeasurementComment}
which we instantaneously and infinitely weakly measure on $\mathcal{B}$ at time $t = x$, which is additionally evolving according to some free hamiltonian $K(t)$. We do this by introducing a collection $\mathcal{A}$ of $n$ meters, labelled by $r = 1, 2, \ldots, n$. The total hamiltonian is given by
\begin{eqnarray}\label{eq:cmham}
	H(t) &=& K(t)\otimes \id_\mathcal{A} + H_I(t),
\end{eqnarray}
where
\begin{eqnarray}
	H_I(t) &=& \sqrt{\epsilon} \sum_{r=1}^n \delta(t-r\epsilon) \left(iR(r\epsilon)\otimes a^\dag_{r\epsilon} + \textrm{h.c.}\right).\nonumber
\end{eqnarray}
We are interested in the limit where $n\rightarrow \infty$ and $\epsilon\rightarrow 0$ with $n\epsilon = L$ fixed. (Indeed, for finite $n$,
this approach includes the scheme of Ref.\ \cite{iblisdir:2007a} when locally representing the meters in terms of coherent states).
The choice of the coefficient $\sqrt{\epsilon}$ in the definition of $H_I$ is motivated by general considerations; any other scaling would lead to trivial dynamics, thanks to the quantum Zeno effort, or to the situation where the meters and system do not interact \cite{caves:1987a}. 

We now supply an interpretation of the cMPS of \cite{verstraete:2010a} based on the sequential preparation prescription of \cite{Schon} and a continuous measurement scenario \cite{sqcmnote}. 
It is straightforward to integrate the Schr\"odinger equation for Eq.~(\ref{eq:cmham}): $U(L)  = \mathcal{T}e^{-i\int_0^L H(s)ds}$, where $\mathcal{T}$ denotes time-ordering, so that ${U}(L) = V(L, L-\epsilon)W(L-\epsilon)V(L-\epsilon, L-2\epsilon)W(L-2\epsilon) \cdots W(\epsilon)V(\epsilon, 0),$ where 
$V(b,a) = \mathcal{T}e^{-i\int_a^b K(s)ds}$ and $W(r\epsilon) = 
e^{\sqrt{\epsilon}\left(R(r\epsilon)\otimes a^\dag_{r\epsilon} - R^\dag(r\epsilon)\otimes a_{r\epsilon}\right)}$. 
By making the standard definition of the discretised {\it quantum field operator} as 
$\Psi(r\epsilon) = a_{r\epsilon}/\sqrt{\epsilon}$ we can, in the limit $\epsilon\rightarrow 0$, describe this evolution via 
\begin{eqnarray}\label{eq:uprop}
	U(L) = \mathcal{T}e^{-i\int_0^L \left(K(s)\otimes \id_\mathcal{A} + iR(s)\otimes \Psi^\dag(s) - iR^\dag(s)\otimes \Psi(s)\right)ds}.
\end{eqnarray}
In our continuum limit the collection $\mathcal{A}$ of the meters may be regarded as a geometrically one-dimensional bosonic quantum field. The auxiliary system $\mathcal{B}$ may be regarded as a geometrically zero-dimensional bosonic quantum field. 

We now learn that the evolution Eq.~(\ref{eq:uprop}) prepares cMPS. Indeed, if we initialise the meters $\mathcal{A}$ in the 
vacuum state vector $|\Omega\rangle_{\cal A}$ of the quantum field we can, for each $r$, exploit the Baker-Hausdorff formula to first order to arrive at the identity 
\begin{eqnarray*}
	&&e^{\sqrt{\epsilon}(R(r \epsilon) \otimes a^\dag_{r \epsilon} - R^\dag(r \epsilon)\otimes 
	a_{r \epsilon})}\id_\mathcal{B}\otimes |0\rangle_\mathcal{A}\\ &=&  
	e^{-\frac{\epsilon}{2}R^\dag(r \epsilon) R(r \epsilon)} e^{\sqrt{\epsilon}R(r \epsilon)\otimes a^\dag_{r \epsilon}}
	\id_\mathcal{B} \otimes |0\rangle_\mathcal{A},\nonumber
\end{eqnarray*}
valid to $O(\epsilon)$, to rewrite the limit Eq.~(\ref{eq:uprop}) for $U(L)(|\Omega\rangle_{\mathcal{A}}\otimes \id_{\mathcal{B}})$ as
\begin{equation*}
	 \mathcal{T}e^{\int_0^L \left(Q(s)\otimes \id_\mathcal{A} 
	 + R(s)\otimes \Psi^\dag(s) \right)ds}(|\Omega\rangle_{\mathcal{A}}\otimes \id_{\mathcal{B}}),
\end{equation*}
where $Q(s) = -iK(s) -\frac12R^\dag(s) R(s)$. This is---up to a trace over the auxiliary degree of freedom---identical 
to the definition of a cMPS \cite{verstraete:2010a}. 

Thus, a cMPS is a quantum state of the quantum 
field $\mathcal{A}$ preparable via a continuum measurement scenario: i.e., we initialise the quantum field $\mathcal{A}$ in 
{\it some} (known) prespecified quantum state 
$\omega_{\mathcal{A}} = |\Omega\rangle_{\mathcal{A}}\langle \Omega|$ and adjoin an auxiliary zero-dimensional quantum field $\mathcal{B}$ initialised in some (possibly mixed) fiducial state $\rho$. We then interact $\mathcal{A}$ and $\mathcal{B}$ according to the \emph{manifestly unitary} continuous measurement dynamics $U(L)$. We then discard the auxiliary system by tracing it out to obtain the quantum state
\begin{equation*}
	\sigma_{\mathcal{A}} = \tr_{\mathcal{B}}\left[U(L) (\omega_{\mathcal{A}}\otimes \rho_{\mathcal{B}}) U^\dag(L)\right].
\end{equation*}

We pause here to emphasise an important point: there is no reason for the initial state $\omega_{\mathcal{A}}$ of the quantum field $\mathcal{A}$ to be the vacuum. All we require is that initial field state arises from the continuum limit of a \emph{product state} $\omega_r \otimes  \cdots\otimes \omega_1$ of the meters $r=1, \ldots, n$. Thus, crucially, we can allow for initial field states with a high density of bosons and superpositions of bosons. This will be important in a variety of contexts, particularly those pertaining to dense systems with nonlinear interactions, where a cMPS defined using the vacuum  will be insufficient. Physically this can be explained as follows: the interaction $H_I$ between $\mathcal{A}$ and $\mathcal{B}$ can only transport $\epsilon$ bosons to $\mathcal{A}$ in a time $\epsilon$. Thus the interaction can never achieve superpositions of terms with more than a constant number of bosons per unit length.

Suppose $\sigma_{\mathcal{A}}$ is a cMPS. We now investigate the dynamics of the auxiliary system throughout the continuous 
measurement process by instead tracing out the field $\mathcal{A}$. Again, we first consider the discrete setting and take the continuum 
limit. We set $\rho(0)=\rho_{\mathcal{B}}$ and  
\begin{equation*}
	\rho(r\epsilon) = \tr_{\mathcal{A}}\left[U(r\epsilon) 
	( \omega_{\mathcal{A}}\otimes \rho(0)) U^\dag(r\epsilon)\right], 
\end{equation*}
where now $U(r\epsilon)  = \mathcal{T}e^{-i\int_0^{r\epsilon} H(s)ds} $. We then consider $\frac{1}{\epsilon}(\rho((r+1)\epsilon)- \rho(r\epsilon))$ and
expand $U(r\epsilon)$ to second order (just as it is done when
describing dissipative quantum systems in the weak coupling limit to arrive at dynamical semi-groups \cite{Alicki}). In our case this is necessary because the field operators contain a factor of $\sqrt{\epsilon}$. 
In the continuum limit
where $n\rightarrow\infty$ and 
$\varepsilon\rightarrow 0$, we arrive at a differential equation for $\rho(x) $ with $x\in[0,L]$,
\begin{eqnarray}\label{eq:lindblad}
	\frac{d\rho}{dx} &=& -i[K, \rho] + \frac12\left(\langle {\Psi^\dag}^2\rangle [R, [R, \rho]]  + 
	\textrm{h.c.}\right) \\ 
	&-& \frac12
	\left(\langle {\Psi^\dag\Psi}\rangle R[R^\dag, \rho] + \langle {\Psi \Psi^\dag}\rangle[\rho, R^\dag]R + \textrm{h.c.}\right),\nonumber\\
	&=& -i[K,\rho] -  \frac12\sum_{j=1}^4 \left(
	{[} M_j^\dagger M_j, \rho{]}_+  - 2 M_j \rho M_j^\dagger\nonumber
	\right),
\end{eqnarray}
where in this expression all operators are evaluated at position $x$, i.e., $K= K(x)$, $\Psi = \Psi(x)$, etc.  
This is an example of a generator of {\it dissipative dynamics} which are manifestly completely positive, and up to dependence of
$x$ being of the form of a {\it Lindblad generator} \cite{Alicki}. Here, the Lindblad operators are identified as
$M_1= i a R - b R^\dagger$, $M_2= i a R + b R^\dagger$, $M_3 = c R^\dagger$, and $M_4=d R$,
where $a^2= \langle {(\Psi^\dag)^2}\rangle/2$, $b^2= \langle {(\Psi)^2}\rangle/2$, $c^2=\langle {\Psi^\dag\Psi}\rangle$,
and $d^2=\langle {\Psi \Psi^\dag}\rangle$. 
We write this equation as $\rho'(x) = \mathcal{L}_x(\rho(x))$. 

We now describe a key feature of cMPS, shared by MPS and PEPS \cite{blogpost}, namely, their \emph{holographic property}. What we mean here is that is the dynamics of a one-dimensional quantum field theory $\mathcal{A}$ in a cMPS is described by the dissipative dynamics of the \emph{boundary} zero-dimensional field theory $\mathcal{B}$ \emph{alone}. This is strongly reminiscent of the \emph{holographic principle} \cite{bousso:2002a}. 
The first step towards the holographic property is to show how expectation values of field operators may be obtained in terms of the 
dynamics of the auxiliary system $\mathcal{B}$ \emph{alone}. This is easy to establish using the following calculational principles, introduced in \cite{verstraete:2010a}. Let $A$ be any 
observable on ${\mathcal{A}}$  which is some product of the field operators and their derivatives at locations $x_1, \ldots, x_n\in [0,L]$ 
in the continuum. 
The first step is to put the observable into normal order with all field annihilation operators on the right. Now we must calculate $\tr[A\sigma_\mathcal{A}] = \tr_{\mathcal{A}}\left[(A \otimes \id_\mathcal{B} )U(L) (\omega_\mathcal{A} \otimes \rho(0)) U^\dag(L)\right]$. To eliminate the field operators we exploit the formula
\begin{equation*}
	[\Psi(x), U(L)] = -i\int_0^L U(L-s) [\Psi(x), F(s)] U(s) ds,
\end{equation*}
where $F(s) = Q(s) + R(s)\otimes \Psi^\dag(s)$, to commute all field annihilation operators past $U(L)$. 
Temporarily assuming \cite{finitefilling} that 
$\Psi(x)|\Omega\rangle_{\mathcal{A}} = 0$ we thus learn that 
$\Psi(x)U(L)|\Omega\rangle_{\mathcal{A}}\otimes \id = U(L-x)R(x)U(x)|\Omega\rangle_{\mathcal{A}}\otimes \id$. Similarly, to evaluate derivatives of $\Psi$ we follow the same procedure with an additional integration by parts to eliminate the derivative of the delta function. We find
$\Psi'(x)U(L)|\Omega\rangle_{\mathcal{A}}\otimes \id_{\mathcal{B}} = \frac{d}{dx}(U(L-x)R(x)U(x))|\Omega\rangle_{\mathcal{A}}\otimes \id_{\mathcal{B}} = 
U(L-x)(-[Q(x), R(x)] + R'(x))U(x)|\Omega\rangle_{\mathcal{A}}\otimes \id_{\mathcal{B}}$. 
Higher derivatives may be evaluating using these methods, but are considerably more tedious. To proceed, it is expedient to employ the \emph{Jamio{\l}kowksi isomorphism} 
where operators $M = \sum_{jk} m_{j,k}|j\rangle\langle k|$ are identified with quantum state 
vectors via $|M\rangle = \sum_{jk} m_{j,k}|j,k\rangle$ \cite{Jami}. 
E.g., in this way, Eq.~(\ref{eq:lindblad}), with $\langle {\Psi^\dag}^2\rangle = \langle {\Psi}^2\rangle = \langle \Psi^\dag\Psi\rangle = 0$, and $\langle \Psi\Psi^\dag\rangle = 1$ is written as
\begin{equation}\label{LB}
	\frac{d}{dx}|\rho(x)\rangle = L(x)|\rho(x)\rangle, 
\end{equation}
where the {\it Liouvillian} is now
$L(x) = -iK\otimes \id + i\id\otimes K^T  -\frac12 \left(R^\dag R\otimes \id - 2R\otimes\overline{R} + \id\otimes R^T\overline{R} \right)$.
Thus, to evaluate a correlation function $\tr[A\sigma_\mathcal{A}]$ we simply have to integrate 
Eq.\ (\ref{LB})
 with \emph{additional insertions} of the operators $R(x)\otimes \id$ at the locations of $\Psi$, $\id\otimes \overline{R}(x)$ at the locations of $\Psi^\dag$, $(-[Q(x),R(x)]+R'(x))\otimes \id$ at the locations of $\Psi'$, etc. We have thus completely eliminated the field $\mathcal{A}$. 
 
 To complete the derivation of the holographic property we need to show how to differentiate a one-parameter family of a cMPS $\sigma_{\mathcal{A}}(t)$, where $t$ is (real or imaginary) time. This is easily achieved using $\frac{d}{dt}U(L,t) = \int_0^L U(L-s) (\frac{\partial}{\partial t} F(s,t)) U(s) ds$, which implies that $\tr(A\sigma'_{\mathcal{A}}(t))$ can also be evaluated in terms (an integral) of the solution of the Lindblad equation with an insertion of $(\frac{\partial}{\partial t} F(s,t))\otimes \id + \id\otimes (\frac{\partial}{\partial t} F^T(s,t))$ at $s=t$. 
 
There is a particularly convenient way to package the procedure we have just described: we define the \emph{generating functional} $Z[J]$,
\begin{eqnarray}\label{eq:genfunct}
	Z[J] = \langle\Omega|\mathcal{T}\exp\left[\int_0^L dx\, L_x + J(\lambda(x), \mu(x))\right] |\Omega\rangle, 
\end{eqnarray}
where $J(\lambda(x), \mu(x)) = \lambda(x)R(x)\otimes \id + \overline{\lambda}(x)\id\otimes \overline{R}(x) + \mu(x)(-[Q(x),R(x)]+R'(x))\otimes \id + \overline{\mu}(x)\id\otimes (-[\overline{Q}(x),\overline{R}(x)]+\overline{R}'(x))$. 
Using $Z[\lambda(x), \mu(x)]$ we can obtain the expectation value of any 
field operator $\Psi$ or its derivative $\Psi'$ via \emph{functional derivatives} with respect to $\lambda(x)$ and $\mu(x)$ (see, e.g., chapter $10$ of Ref.\ \cite{peskin:1995a} for a brief introduction to functional derivatives), e.g., 
\begin{equation}
	\left\langle\Psi^\dag(z)\Psi'(y)\Psi(x) \right\rangle  =  \frac{\delta^3 Z[J]}{\delta \lambda(x) \delta \mu(y)\delta \overline{\lambda}(z)} \bigg|_{\lambda, \mu = 0}.
\end{equation}
Now we have introduced the generating functional $Z[J]$ it is straightforward to generalise the cMPS construction to arbitrary zero-dimensional field theories (i.e., $D\rightarrow \infty$) via the path-integral prescription: we promote $K$ and $R$ to functions of field operators and evaluate the time-ordered integral Eq.~(\ref{eq:genfunct}) via the standard path-integral prescription. Since the boundary field theory is a 
non-equilibrium theory it is convenient to employ the {\it Keldysh formalism} to evaluate the generating functional. 
In this way we can easily understand some general properties of cMPS. 

For finite $D$, and in the translationally invariant setting where $R$ and $K$ do not depend on $x$, it is particularly
transparent to see that generically, all correlation functions decay exponentially. Let us assume that the Liouvillian $L$ 
generating  \emph{Markovian dynamics} has
a unique zero eigenvalue and the real part of any other eigenvalue is bounded from above by $-\Delta$, 
$\Delta>0$ being a {\it gap}. Since for $\Psi(x)|\Omega\rangle_{\cal A}=0$ we have that
$\langle \Psi^\dagger(x_1) \Psi(x_2)\rangle = 
\tr_{\mathcal{A}}[ U(L-x_2) R U(x_2) (\omega_\mathcal{A} \otimes \rho_{\cal B})U^\dagger (x_1)R^\dagger  U^\dagger (L-x_1)]$,
we can apply the above rule for integrating the master equation, using techniques to compute two-point correlation functions
for dynamical semi-groups \cite{Alicki}, to see that 
there exists a suitable $c>0$ with
$|\langle \Psi^\dagger(x_1) \Psi(x_2)\rangle|\leq c e^{-\Delta(x_2-x_1)}$. Similarly, all other  
\emph{spatial} correlators of our original field $\mathcal{A}$ are clustering.

\begin{figure}
	\includegraphics{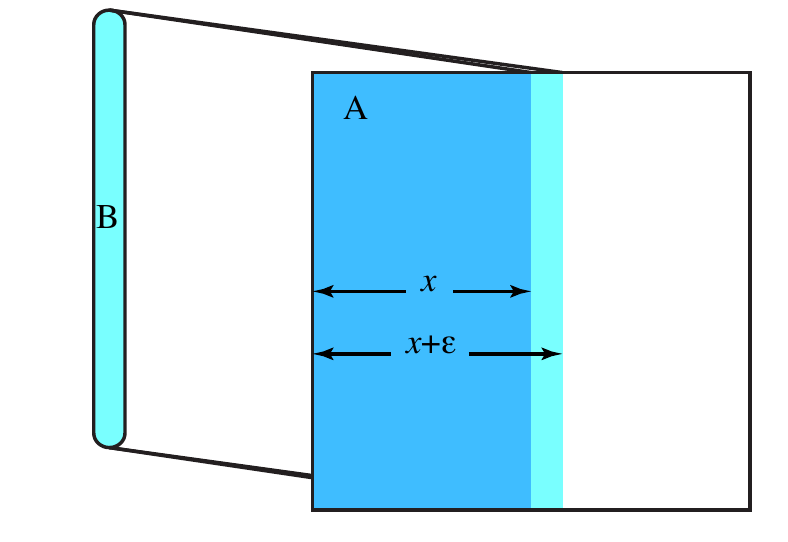}
	\caption{Here we illustrate the physical process underlying the construction of a $2+1$-dimensional generalisation. The system $\mathcal{A}$ is initialised in the vacuum state $|\Omega\rangle$ and then infinitesimally thin vertical strips at horizontal location $x$ are sequentially interacted with $\mathcal{B}$ at time $t=x$.}\label{fig:2dfcs}
\end{figure}

The perspective offered here from the viewpoint of continuous measurement and the holographic principle allows us to easily generalise the cMPS ansatz class to field theories with a larger geometric dimension. We describe this generalisation in terms of a $2+1$-dimensional bosonic theory, because the generalisation to higher dimensions offers no new complications (such a generalisation was anticipated in \cite{verstraete:2010a}). Intuitively our construction may be described as follows. Suppose we have a two (spatial) dimensional bosonic field theory $\mathcal{A}$ with field operator $\Psi(x,y)$, $x,y\in [0, L]$. Motivated by our $1+1$-dimensional construction we propose to introduce an \emph{auxiliary} $1+1$-dimensional field theory $\mathcal{B}$ described by a tuple of field operators $\Phi_{\alpha}(y)$, $\alpha = 1, 2, \dots, D$, (which may be transforming as a spinor, vector, or matrix etc.), which we think of as living vertically ``at the boundary'' of $\mathcal{A}$. To prepare a quantum state for $\mathcal{A}$ we work entirely analogously as to before: we initialise $\mathcal{A}$ in some fiducial state, say the vacuum $|\Omega\rangle$. We then interact an ``infinitesimally thin'' vertical strip of $\mathcal{A}$ and $\mathcal{B}$ according to some \emph{spatially local} interaction $\mathcal{R}(0)$, where 
\begin{equation}
	\mathcal{R}(x) = i\int_0^L dy\, R_x(\Phi_\alpha(y), \Phi'_\alpha(y))\otimes \Psi^\dag(x,y)+\textrm{h.c.},
\end{equation}
and where $R_x(\Phi_\alpha(y), \Phi'_\alpha(y))$, which may depend on the position $x$, is some polynomial in the field operators $\Phi_\alpha(y)$, their derivatives, and their adjoints at location $y$. We then proceed by interacting infinitesimal vertical strips at locations $x$ of $\mathcal{A}$ and $\mathcal{B}$ sequentially at times $t=x$. Interleaved between each interaction between the strip at $x$ and $\mathcal{B}$ we also evolve $\mathcal{B}$ according to free dynamics generated by some 
\begin{equation}
	\mathcal{K}(x) = \int_0^L dy\, K_x(\Phi_\alpha(y), \Phi_\alpha'(y)),
\end{equation}
where $K_x(\Phi_\alpha(y), \Phi'_\alpha(y))$ is some hermitian operator written in terms of a polynomial in the field operators $\Phi_\alpha(y)$, their derivatives, and their adjoints at location $y$. This unitary process, illustrated in Fig.~(\ref{fig:2dfcs}), is described by the propagator
\begin{equation}
	U(L,L) = \mathcal{T}e^{-i\int_0^Ldx\, \mathcal{K}(x) + (i\mathcal{R}(x) +\textrm{h.c.})}.
\end{equation}
We can now apply, \emph{without modification}, the analysis derived earlier to describe the holographic property of the states generated by $U(L,L)$. A central role is again played by the Lindblad equation
\begin{multline}
	\frac{d\rho}{dx} = -i[\mathcal{K}(x), \rho] \\ -\frac12 \int_0^{L} dy\, [R_x^\dag(\Phi_{\alpha}(y)), R_x(\Phi_{\alpha}(y))\rho] + \textrm{h.c.}.
\end{multline} 
(This is the equation for the case where the initial state of $\mathcal{A}$ is the vacuum. The case of finite filling is more involved but is of a similar form to the $1+1$-dimensional case.)
This describes a local dissipative field theory for $\mathcal{B}$; expectation values of physical operators may be recovered by integrating this equation with the appropriate insertions. Note that since the entire process is unitary all expectation values arising from integration of this equation are \emph{physical}.

While it is clear that, in the limit $L\rightarrow\infty$, when $K_x$ and $R_x$ are translation invariant the states arising from this construction will be translation invariant \cite{topology} it is far less obvious what conditions must be imposed on $K_x$ and $R_x$ in order that the resulting state is \emph{rotation invariant}. This is a more subtle problem and there are two points of view one can take here: the first is that we simply ignore the rotation invariance issue and when we use 2D generalisation as a variational class we assume that the optimal state will inherit the symmetries of the hamiltonian. The second point of view is that we should explicitly characterise the rotation invariant instances and use the resulting subclass as a variational class for rotation-invariant systems, as this ought to be a simpler variational problem. In both cases we need to at least satisfy ourselves that rotation invariant 2D generalisations \emph{exist}. In order to argue this we suppose that the dissipative boundary theory for $\mathcal{B}$ has a unique \emph{spacetime} rotation invariant fixed point, which is equivalent to translation invariance and the condition that $R_x(\Phi_\alpha(y))$ transforms as a \emph{scalar} under spacetime rotations. Then all we need is that the free dynamics of $\mathcal{B}$ generated by $\mathcal{K}$ is also \emph{spacetime} rotation invariant. This is a nontrivial condition and requires that the field $\Phi_\alpha$ transforms in a representation of $\textsl{SO}(2)$, i.e., as a vector. These two conditions suffice for the resulting 2D generalisation to be rotation invariant. In the same way: if we want that the 2D generalisation is \emph{conformally invariant} then we require that the boundary theory for $\mathcal{B}$ is also conformally invariant \cite{inpreparation}.

In this work we discussed an interpretation of a recently introduced variational class, continuous matrix product states, for bosonic quantum fields.  We
have explained how this class arises naturally from the procedure
of continuous measurement, and used this observation
to explain the key physical properties of cMPS, including
the clustering of correlations. We also discovered a fundamental
holographic property possessed by cMPS, namely that the dynamics
of a quantum field in an cMPS can be completely understood
in terms of a (dissipative) boundary field theory of
one dimension lower. Finally, we have expressed the definition
of an cMPS in purely field-theoretic terms which facilitates
generalisations to higher-dimensional systems and
the use of perturbation theory.

We have also pointed out that the zero-dimensional boundary field corresponds to the internal degrees of freedom of an atom in cavity QED experiments; this opens up the possibility of simulating quantum field theories with simple dissipative dynamics. 

A great many future directions present themselves at this
point: one can explore the utility of cMPS as a variational
ansatz for numerical calculations. Additionally, the extension
of the theory of cMPS for fermionic fields and gauge fields is possible \cite{inpreparation}. Finally, the relationship of the
regulator that cMPS provides with other standard renormalisation
prescriptions (e.g., dimensional regularisation) remains
to be elucidated.

\emph{Acknowledgements}. F.V.\ thanks I.\ Cirac, M.\ Fannes, and R.\ Werner for very useful discussions, and the SFB projects FoQuS and ViCoM and European projects Quevadis and ERC grant QUERG for funding. J.E.\ thanks the Qessence, Minos, Compas, and EURYI projects.


\begin{thebibliography}{99}
\bibitem{verstraete:2005a}
	F.\ Verstraete and J.I.\ Cirac, Phys.\ Rev.\ B {\bf 73}, 094423 (2006). %

\bibitem{osborne:2005d}
	T.J.\ Osborne, Phys.\ Rev.\ Lett.\ {\bf 97}, 157202 (2006). %

\bibitem{hastings:2007a}
	M.B.\ Hastings, J.\ Stat.\ Mech.\ P08024 (2007). %

\bibitem{area}
	J.\ Eisert, M.\ Cramer, and M.B.\ Plenio, Rev.\ Mod.\ Phys.\ {\bf 82}, 277 (2010). %

\bibitem{fannes:1991a}
	M.\ Fannes, B.\ Nachtergaele, and R.F.\ Werner, J.\  Phys.\ A {\bf 24}, L185 (1991). %

\bibitem{Wilson} K.\ G.\ Wilson, Rev. Modern Phys. {\bf 47}, 773--840 (1975). %

\bibitem{White} S.\ R.\ White, Phys. Rev. Lett. {\bf 69}, 2863--2866 (1992)  %

\bibitem{schollwoeck:2005a}	
	U.\ Schollw{\"o}ck, Rev.\ Mod.\ Phys.\ {\bf 77}, 259 (2005).%

\bibitem{richter:1996}
	S.\ Richter and R.F.\ Werner, J.\ Stat.\ Phys.\ {\bf 82}, 963 (1996).

\bibitem{verstraete:2004a}
	F.\ Verstraete and J.I.\ Cirac, cond-mat/0407066. %

\bibitem{jordan:2008a}
	J.\ Jordan, R.\ Or\'us, G.\ Vidal, F.\ Verstraete, and J.\ I.\ Cirac, Phys. Rev. Lett. {\bf 101}, 250602 (2008).%

\bibitem{iblisdir:2007a}
	S.\ Iblisdir, R.\ Orus, and J.I.\ Latorre, Phys.\ Rev.\ B {\bf 75}, 104305 (2007). %

\bibitem{muth:2009a}
	D.\ Muth, B.\ Schmidt, and M.\ Fleischhauer, arXiv:0910.1749. %

\bibitem{weir:2010a} 
	D.\ J.\ Weir, arXiv:1003.0698. %

\bibitem{verstraete:2010a}
	F.\ Verstraete and J.I.\ Cirac, arXiv:1002.1824. %

\bibitem{maruyama:2010a}
	I.\ Maruyama and H.\ Katsura, arXiv:1003.5463. %

\bibitem{caves:1987a}	
	C.M.\ Caves and G.J.\ Milburn, Phys.\ Rev.\ A {\bf 36}, 5543 (1987). %

\bibitem{Feynman} R. Feynman, Proc. Int. Workshop on Variational Calculus in Quantum Field Theory, Wangerooge, West Germany (World Scientific, Singapore, 1987). %

\bibitem{Zoller} P. Zoller, M. Marte, D.F. Walls, Phys. Rev. A {\bf  35}, 198 (1987); C.W. Gardiner and P. Zoller, Quantum Noise (Springer-Verlag, Berlin, 2004). %

\bibitem{Kimble} H.J. Kimble,  Phys. Scripta, {\bf T76}, 127-137  (1998). %

\bibitem{Rempe} M. Hijlkema {\it et al.}, Nat. Phys. {\bf 3},  253  (2007). %

\bibitem{Schon} C. Schoen et al., Phys. Rev. Lett. {\bf 95}, 110503 (2005). %

\bibitem{Lieb-Liniger}  E. H. Lieb and W. Liniger, Phys. Rev. {\bf 130}, 1605 (1963). %

\bibitem{Garrahan} J. P. Garrahan  and I. Lesanovsky, Phys. Rev. Lett. {\bf 104}, 160601 (2010). %

\bibitem{vonneumann:1996a}
	J.\ von Neumann, {\it Mathematical foundations of quantum mechanics} (Princeton University Press, Princeton, 1996). %

\bibitem{POVMComment}
	A POVM is the most general kind of quantum measurement: it is defined by a set $\{E_j\}_j^{N}$ of \emph{positive} operators satisfying $\sum_{j=1}^N E_j = \id$. If the quantum system is in state $\rho$ the probability $p_j$ of outcome $j$ is given by $p_j = \tr(E_j\rho)$. %

\bibitem{MeasurementComment}
	In order to actually measure an arbitrary non-Hermitian matrix $R$ on our quantum system we must first develop a modification of von Neumann's prescription: we suppose the meter is now a \emph{harmonic oscillator} initially in the ground state, which is a coherent state. Supposing the system is in $|\phi\rangle$ the effect of measuring $R$ can be effected by coupling the system and meter according to the interaction
\begin{equation}
	H_I = iR\otimes a^\dag - iR^\dag\otimes a
\end{equation} 
and evolving for a time $t$. If we temporarily assume that $R$ is a \emph{normal} matrix, $[R,R^\dag]=0$, then the resulting state vector of the system and meter is given by $e^{-itH_I}|\Psi\rangle|0\rangle = \sum_{j=1}^D \psi_j |r_j\rangle |r_jt\rangle,$ where $R = \sum_{j=1}^D r_j|r_j\rangle\langle r_j|$ is the eigenvalue decomposition of $R$, $r_j$ are the \emph{complex} eigenvalues of $R$, and $|\alpha\rangle = D(a,\alpha)|0\rangle$ is a coherent state vector centred on the phase space point $\alpha\in \mathbb{C}$. If we increase $t$ we can arrange for $|r_jt\rangle$ to become arbitrarily well distinguished by a physical measurement of the meter alone (eight-port homodyne detection \cite{walker:1987a}). Note that this procedure is \emph{physical} even though the measurement of a non-Hermitian operator does defy immediate physical interpretation: it is a result of unitary evolution and standard quantum-mechanical measurement. %

\bibitem{sqcmnote}
	This is related to an interpretation supplied in \cite{verstraete:2010a}.
\bibitem{Alicki}
	R.\ Alicki and K.\ Lendi, {\it Quantum dynamical semigroups and applications} (Springer, Berlin, 1997). %

\bibitem{blogpost}
		T.J.\ Osborne, {\it Holographic quantum states},  {http:// tjoresearchnotes.wordpress.com/2010/05/07/holographic-quantum-states/}. %
	
\bibitem{bousso:2002a}
	R.\ Bousso, Rev.\ Mod.\ Phys.\ {\bf 74}, 825 (2002). %

\bibitem{finitefilling}
	The case of finite filling is considerably more involved, and we do not reproduce all of the required calculational steps as there are no new tools required beyond those described in the procedure outlined here.
	 
\bibitem{topology}
	For the limit to make sense we must also ensure that the correlations are \emph{clustering} so that the field isn't affected by topology of the system. 
	
\bibitem{walker:1987a}
	N.G.\ Walker, J.\ Mod.\ Opt.\ {\bf  34}, 15 (1987). 

\bibitem{Jami}
	Multiplication by an operator on the left 
	(respectively, right) is carried out using the identification 
	$AM \leftrightarrow A\otimes \id|M\rangle = |AM\rangle$ (respectively, 
	$MA \leftrightarrow \id\otimes A^T|M\rangle = |MA\rangle$).
	
\bibitem{peskin:1995a}
	M.E.\ Peskin and D.V.\ Schroeder, {\it An introduction to quantum field theory} (Addison Wesley, Reading MA, 1995).	
	
\bibitem{inpreparation}
	{\it In preparation}.
\end{thebibliography}
\end{document}